\documentclass[sigconf, screen]{acmart}
\AtBeginDocument{%
  \providecommand\BibTeX{{%
    \normalfont B\kern-0.5em{\scshape i\kern-0.25em b}\kern-0.8em\TeX}}}

\copyrightyear{2024}
\acmYear{2024}
\setcopyright{acmlicensed}\acmConference[OARS 2024]{Proceedings of the 4th International Workshop on Online and Adaptive Recommender Systems}{October 25, 2024}{Boise, ID, USA}
\acmBooktitle{Proceedings of the 4th International Workshop on Online and Adaptive Recommender Systems (OARS 2024), Held in conjunction with CIKM-2024, October 25, 2024, Boise, ID, USA}
\acmDOI{10.1145/3627673.3679083}
\acmISBN{979-8-4007-0436-9/24/10}

\usepackage{balance}
\usepackage{hyperref}
\usepackage{caption}
\usepackage{subcaption}
\usepackage{soul}
\usepackage[inline]{enumitem}
\usepackage[normalem]{ulem}
\usepackage{multirow}
\usepackage{graphicx} % For \resizebox
\usepackage{adjustbox} % For \adjustbox (optional, if more control is needed)
\usepackage{array}
\usepackage{hyperxmp}

\begin{document}

\newcommand{\srijan}[1]{\textcolor{blue}{[srijan: #1]}}
\newcommand{\vivek}[1]{\textcolor{red}{[Vivek: #1]}}

\newcommand{\E}{\mathbf{E}}
\newcommand{\x}{\mathbf{x}}
\newcommand{\y}{\mathbf{y}}

\title{Real-time Event Joining in Practice With Kafka and Flink}

%\author{Anonymous author(s)}
%\affiliation{%
%  \institution{anonymous affiliation(s)}
%}

\author{Srijan Saket$^*$}
\email{srijanskt@gmail.com}
\affiliation{
    \institution{ShareChat}
    \city{Seattle}
    \country{USA}
}

\author{Vivek Chandela$^*$}
\email{vivekchandela@sharechat.co}
\affiliation{
    \institution{ShareChat}
    \city{Bangalore}
    \country{India}
}

\author{Md. Danish Kalim}
\email{danish@sharechat.co}
\affiliation{
    \institution{ShareChat}
    \city{Bangalore}
    \country{India}
}

\renewcommand{\shortauthors}{Saket, Chandela et al.}

\makeatletter
\let\@authorsaddresses\@empty
\makeatother

\begin{abstract}

Historically, machine learning training pipelines have predominantly relied on batch training models, retraining models every few hours. However, industrial practitioners have proved that real-time training can lead to a more adaptive and personalized user experience. The transition from batch to real-time is full of tradeoffs to get the benefits of accuracy and freshness while keeping the costs low and having a predictable, maintainable system.

Our work characterizes migrating to a streaming pipeline for a machine learning model using Apache Kafka and Flink. We demonstrate how to transition from Google Pub/Sub to Kafka to handle incoming real-time events and leverage Flink for streaming joins using RocksDB and checkpointing. We also address challenges such as managing causal dependencies between events, balancing event time versus processing time, and ensuring exactly-once versus at-least-once delivery guarantees, among other issues. Furthermore, we showcase how we improved scalability by using topic partitioning in Kafka, reduced event throughput by \textbf{85\%} through the use of Avro schema and compression, decreased costs by \textbf{40\%}, and implemented a separate pipeline to ensure data correctness. Our findings provide valuable insights into the tradeoffs and complexities of real-time systems, enabling better-informed decisions tailored to specific requirements for building effective streaming systems that enhance user satisfaction.

\end{abstract}




%    <concept>
%        <concept_id>10002951.10003317.10003338</concept_id>
%        <concept_desc>Information systems~Recommender systems</concept_desc>
%        <concept_significance>500</concept_significance>
%    </concept>
%    <concept>
%        <concept_id>10010147.10010169.10010170</concept_id>
%        <concept_desc>Computing methodologies~Machine learning algorithms</concept_desc>
%        <concept_significance>500</concept_significance>
%    </concept>
%    <concept>
%        <concept_id>10002951.10003260.10003282</concept_id>
%        <concept_desc>Information systems~Social networks</concept_desc>
%        <concept_significance>500</concept_significance>
%    </concept>
%    <concept>
%        <concept_id>10010405.10010406.10010407</concept_id>
%        <concept_desc>Applied computing~Business metrics</concept_desc>
%        <concept_significance>300</concept_significance>
%    </concept>

% \ccsdesc[500]{Information systems~Recommender systems}
% % \ccsdesc[300]{Computing methodologies~Machine learning algorithms}
% \ccsdesc[500]{Information systems~Social networks}
% \ccsdesc[300]{Applied computing~Business metrics}


\maketitle
\def\thefootnote{\texorpdfstring{*}{}}\footnotetext{Equal Contribution}

\section{Introduction}
In the modern information technology era, users generate large quantities of data. Platforms collect and process this amassed data to derive meaningful insights that can enhance their system. Social media platforms, in particular, generate enormous volumes of data from various user activities \cite{mtl, mtl2, embedding_evolution}. Short-form video (SFV) platforms are especially prominent, offering an immersive viewing experience that attracts substantial user attention. Due to the nature of their offerings, SFV platforms cater to the passive preferences of users, which are inherently dynamic and frequently changing. The quality of users' lean-back experience (i.e., users passively consume information) relies heavily on the system's ability to capture user feedback and adapt recommendations in real-time using user feedback.

\textbf{Stream processing} is a well-known and extensively studied field. Work in this domain has focused on

\begin{enumerate}

    \item Utilising real-time processing framework like Flink to prepare data from machine learning systems \cite{flink, streaming, toshniwal2014storm, noghabi2017samza, kulkarni2015twitter}.

    \item Employing real-time framework for real-time model updates such as Monolith \cite{monolith}, Solma \cite{solma} and others \cite{streamingsvm, kamburugamuve2018anatomy, guo2021stream}.

\end{enumerate}

This paper aims to use a state-of-the-art stream processing framework to implement practical and industry-scale real-time data systems that will enable real-time model building. We will discuss the challenges in building these systems and strategies to overcome them. Key challenges in building industry-scale real-time data systems include:

\begin{enumerate}

    \item Stream Integration: When we deal with data streams containing features and labels, using a connector to merge them into machine learning systems is essential \cite{gomes2019machine, bifet2023machine}. Sometimes, combining labels from different streams into a single stream for multi-task modeling setups is necessary \cite{lin2015scalable, han2022data}.

    \item Out-of-order Events: Managing delays in data streams generated by events is vital in system configurations. The system should be able to process events regardless of their timing to ensure the accuracy of training data. For instance, creating training data where a user has viewed and liked an item but has not shared it would require assigning a value of 1 to the like label and 0 to the share label.

    \item Load Handling: User engagement with content on the platform over time drives the model's load. User behaviors can change, leading to fluctuations like increased usage in the evening compared to during the day, which highlights the importance of efficiently managing back pressure \cite{9016513}.

    \item Concurrent Updates: With the increased volume of data, updates can occur concurrently for a critical element, such as users or items. The system must address conflicts arising from these simultaneous updates to maintain training stability.

\end{enumerate}

\begin{figure*}[t!]
    \centering
    \begin{subfigure}[t]{0.45\textwidth}
        \centering
        \includegraphics[width=\textwidth]{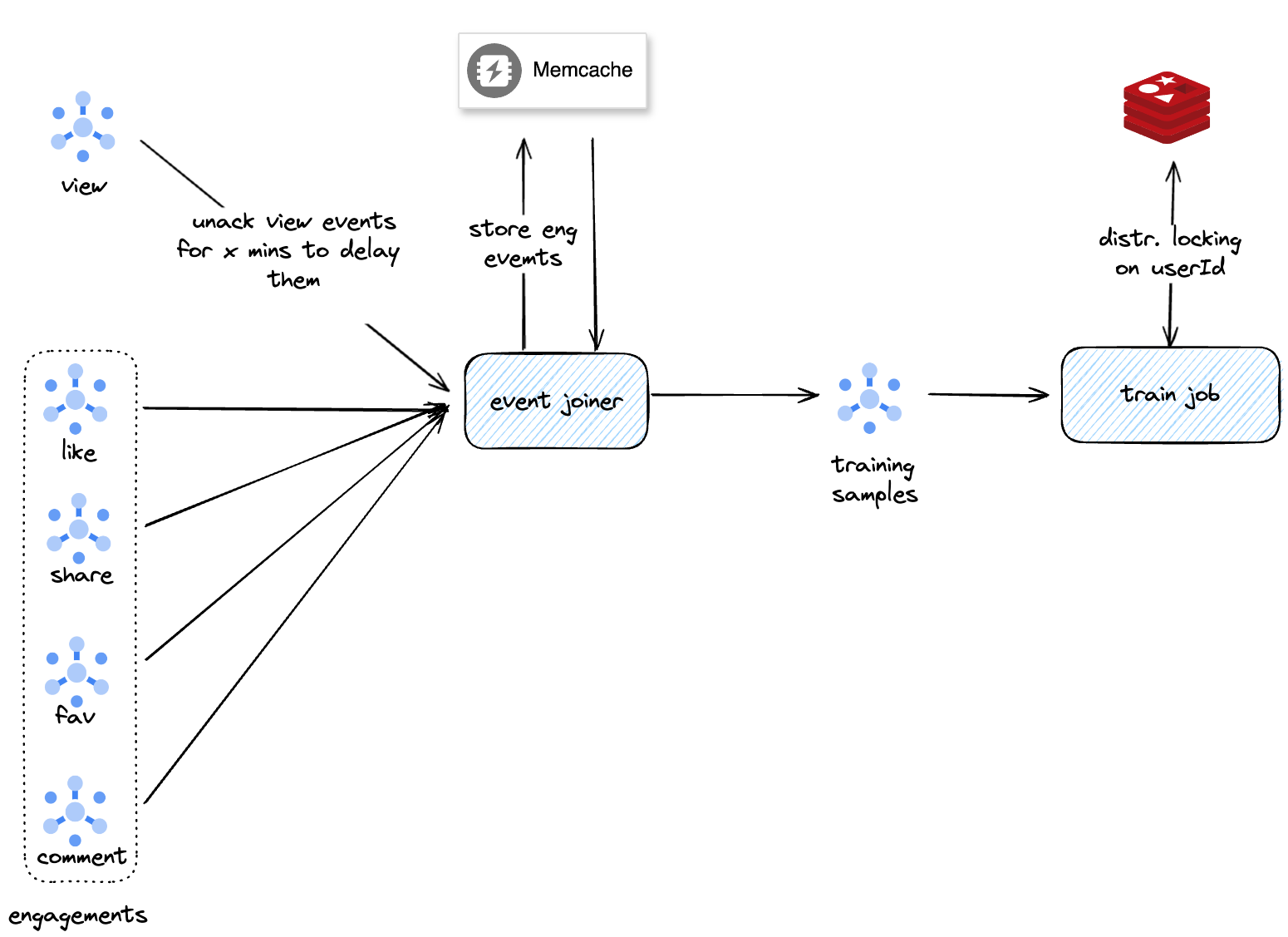}
        \caption{System to join incoming real-time events with labels to prepare training samples; using caching components while maintaining the order of events
        }
        \label{fig:rt_old}
    \end{subfigure}
    \hspace{1mm}
    \begin{subfigure}[t]{0.45\textwidth}
        \centering
        \includegraphics[width=\textwidth]{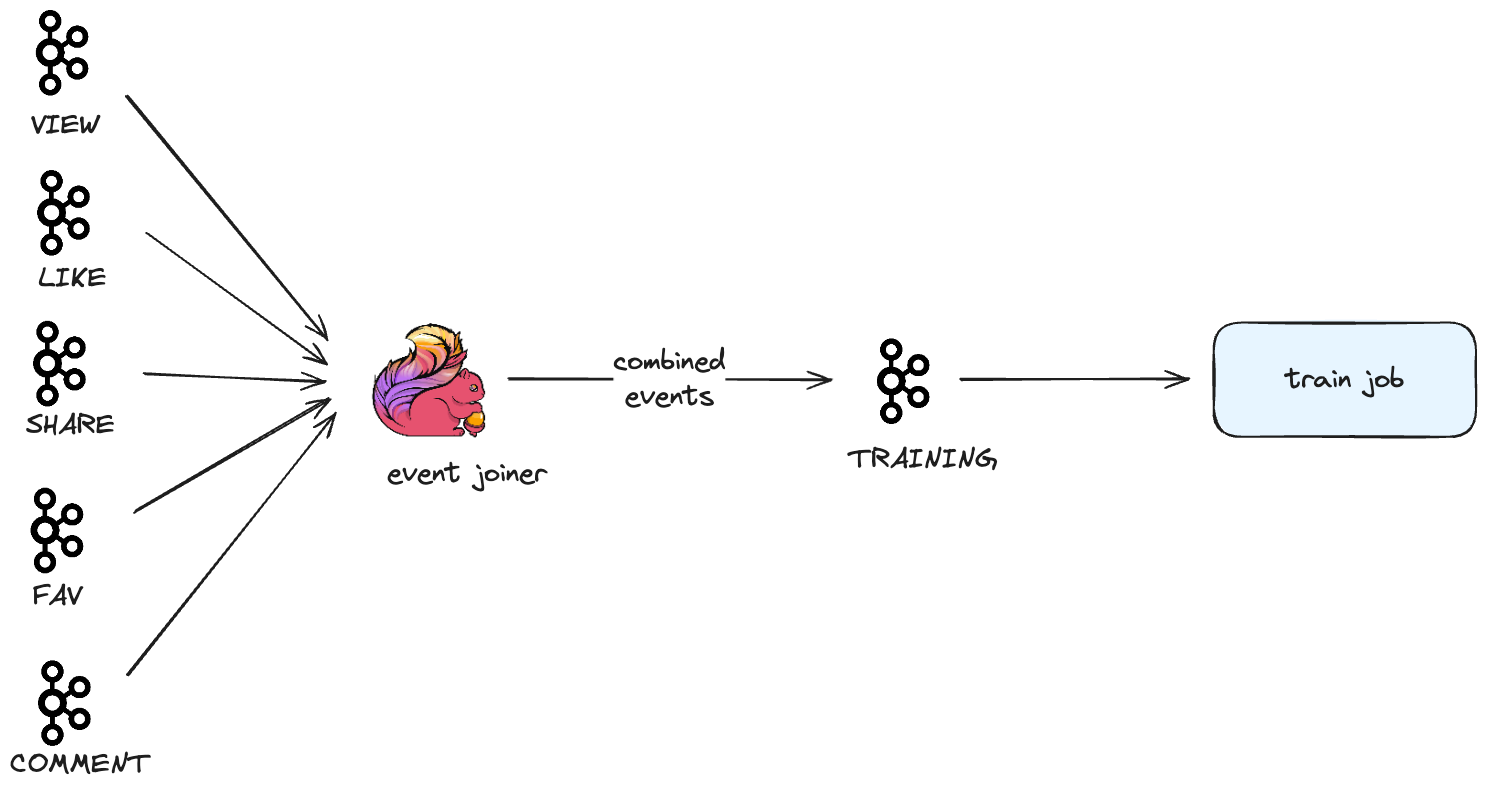}
        \caption{ 
        Modified system to join incoming real-time events with labels to prepare training samples; using Kafka and Flink overcoming the shortcomings of previous approach
        }
        \label{fig:rt_new}
    \end{subfigure}
    \caption{System comparison of Approach 1 vs. Approach 2}
    \vspace{-2mm}
\label{fig:system_comparison}
\end{figure*}

Considering all the above challenges, we built two systems. Both approaches differ in the choice of components, complexity, and maintenance:

\begin{enumerate}

    \item \textbf{Approach 1}: In the setup described in Fig \ref{fig:rt_old}, PubSub \cite{pubsub} is used as the messaging queue. Order of events is preserved (in a time window) by using Memcache \cite{memcache}, also used for real-time label joining and preparing training samples for downstream training jobs. Further, \href{https://redis.io/docs/latest/}{Redis} is used for distributed locking to avoid concurrent updates on the same ID.

    \item \textbf{Approach 2}: The setup in Figure \ref{fig:rt_new} uses Kafka \cite{kafka} as the messaging queue, with its benefits discussed in subsequent sections. By integrating Apache Flink with Kafka, we maintain event order and prevent concurrent updates. Although this approach involves fewer components than the initial setup, we will demonstrate why it is superior through comparison.

\end{enumerate}

\section{Methodology}
\subsection{Data Context \& User Signals}
The study was conducted on \href{https://en.wikipedia.org/wiki/ShareChat}{ShareChat}'s data, which is a multi-lingual social media platform that delivers content in over 18 languages, with a user base exceeding 180 million monthly active users. The application generates two broad categories of events: views and engagements. The system generates view events whenever it shows content to a user. These are high-volume events with a peak throughput of \textit{>100 MB/sec}. Conversely, engagements are user signals comprised of implicit signals (such as video play and skip) and explicit signals (including click, like, and share). However, the event processing pipeline does not distinguish between them. The volume of these events is significantly lower compared to views. We used Field-aware Factorization Machines (FFM) \cite{ffm_paper, juan2017field} to learn 32-dimensional embedding for each signal. Real-time training retrieves the previous state of these embeddings from a NoSQL database and updates them using real-time user signals from messaging queues.

\subsection{Generating Training Samples}
The system generates real-time training samples by joining a view event with its corresponding engagement label, if available \cite{zhang2023leveraging, saket2023formulating}. In this use case, it considers a view with an engagement event as a positive label and a view without engagement as a negative label. For example, if there is a like event corresponding to a view but no share, comment, or favorite, the label for the like will be 1 and 0 for the others. These events are received through different PubSub topics, as we have separate models to predict embeddings for each user signal, as illustrated in Figure \ref{fig:rt_old}.

\begin{figure*}[t!]
    \centering
    \begin{subfigure}[t]{0.35\textwidth}
        \centering
        \includegraphics[width=\textwidth]{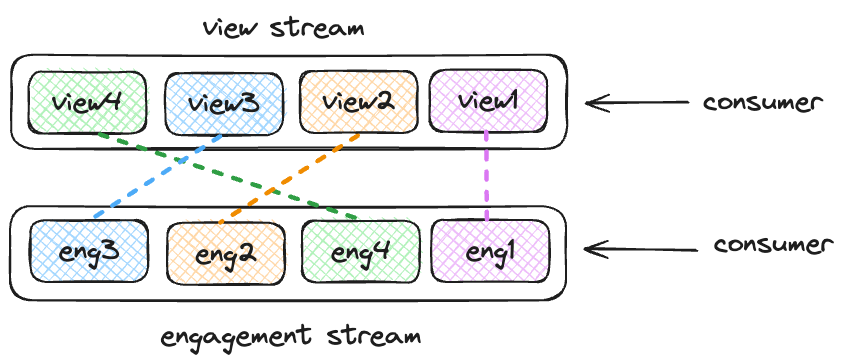}
        \caption{Out-of-order events}
        \label{fig:challenge_1}
    \end{subfigure}
    \hspace{2mm} % Add horizontal space for separation
    \vrule width 0.2pt % Vertical line as a separator
    \hspace{2mm} % Add horizontal space for separation
    \begin{subfigure}[t]{0.35\textwidth}
        \centering
        \includegraphics[width=\textwidth]{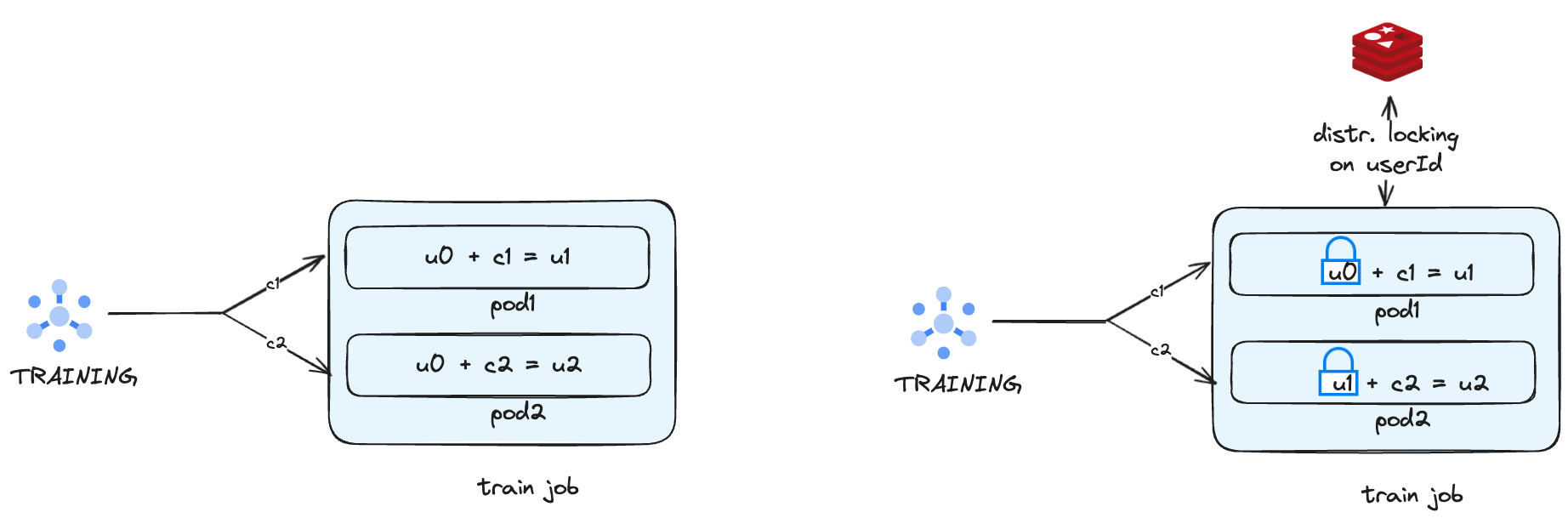}
        \caption{Concurrent updates on an ID}
        \label{fig:challenge_2}
    \end{subfigure}
    \hspace{2mm} % Add horizontal space for separation
    \vrule width 0.2pt % Vertical line as a separator
    \hspace{2mm} % Add horizontal space for separation
    \begin{subfigure}[t]{0.2\textwidth}
        \centering
        \includegraphics[width=\textwidth]{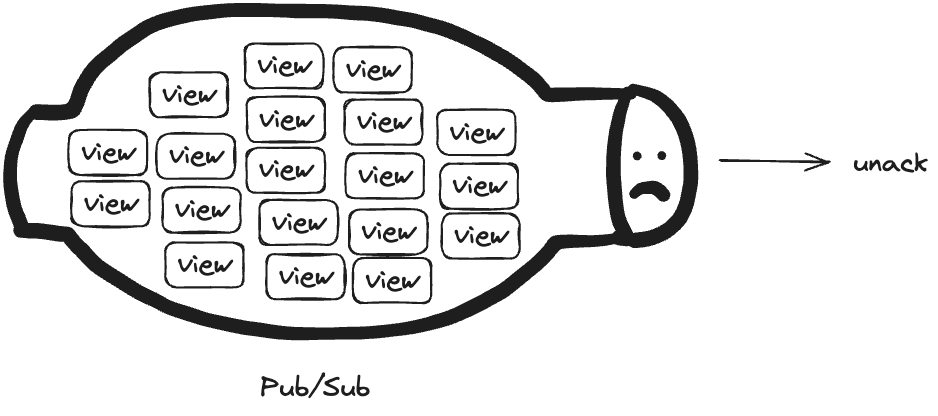}
        \caption{Inflating queue size}
        \label{fig:challenge_3}
    \end{subfigure}
    \caption{Key Challenges in Event Streaming Pipelines}
    \label{fig:key_challenges}
\end{figure*}

\subsection{Key Challenges}
\subsubsection{Handling out-of-order events}
The sequence of events is crucial for making meaningful updates; otherwise, we risk incorporating incorrect information in the future. Since the system processes views and engagements in separate queues, we cannot determine whether a view precedes engagements or vice versa [\ref{fig:challenge_1}]. One approach is to delay view events by a few minutes by holding them in Pub/Sub without acknowledging receipt while temporarily storing engagements in a distributed key-value store like Memcache, with a TTL longer than the delay. However, delaying view events in Pub/Sub by withholding acknowledgment is inefficient.

\subsubsection{Pod Contention}
If two pods update the same user embedding concurrently with different training samples, it can lead to incorrect outcomes due to the lack of order preservation. To prevent this, we need to lock the user ID during updates as described in  [\ref{fig:challenge_2}]. Acquiring a distributed lock on the user ID via Redis prevents concurrent updates, thereby resolving the issue. However, this adds an extra component to the system, increasing costs.

\subsubsection{Inflating Queue Size}
Pub/Sub lacks extensive support for data retention, message replay, and message ordering compared to log-based queues. Once a message is acked in Pub/Sub, it is gone forever. So, the only way to delay view events is by unacking them, which inflates the queue size [\ref{fig:challenge_3}] and our cloud bills.

\subsection{Making ``Better'' System Choices}
Having established the problem statement in the previous sections, let us explore how we can make choices to create a more reliable system with the same outcomes while keeping costs under control.
\subsubsection{Replace PubSub with Kafka}
A good starting point is replacing Google PubSub with Apache Kafka. Kafka is a log-based event-streaming platform that provides message ordering, the ability to replay messages, and longer data retention. Apache Flink has excellent support for Kafka.
Also, self-hosted Kafka proved cheaper than GCP's managed PubSub offering.

\subsubsection{Use Flink}
We then rely on Apache Flink \cite{flink}, an open-source stream processing framework, as the real-time event joiner. Its rich feature set includes internal state management, keyed streams, timers, and many more. Leveraging Flink's support for the state backend, particularly RocksDB, allowed us to eliminate the need for Memcache. Some beneficial properties of Flink in the proposed solution are:
\begin{enumerate}
    \item Graceful Backpressure Handling: Backpressure refers to a system receiving data faster than it can process, often during a temporary load spike. Flink handles backpressure gracefully without any sophisticated mechanism.
    
    \item State: Flink enables operators to retain information across multiple events using state. Flink provides various state primitives like ValueState for single values, MapState for key-value pairs, and many more. A TTL can be assigned to any state as needed.
    
    \item Low-level APIs: Flink supports both high-level and low-level APIs. We chose low-level APIs like \texttt{KeyedProcess\-Function} and \texttt{KeyedCoProcess\-Function} for joining view and engagement events, as they give more control over each event.
    
    \item Checkpointing: The central part of Flink's fault tolerance mechanism is drawing consistent snapshots of all the states in timers and stateful operators. The system can fall back to the latest snapshot in case of failure.
\end{enumerate}

\begin{figure}[!t]
\begin{minipage}[b]{\linewidth}
    \includegraphics[width=\textwidth]{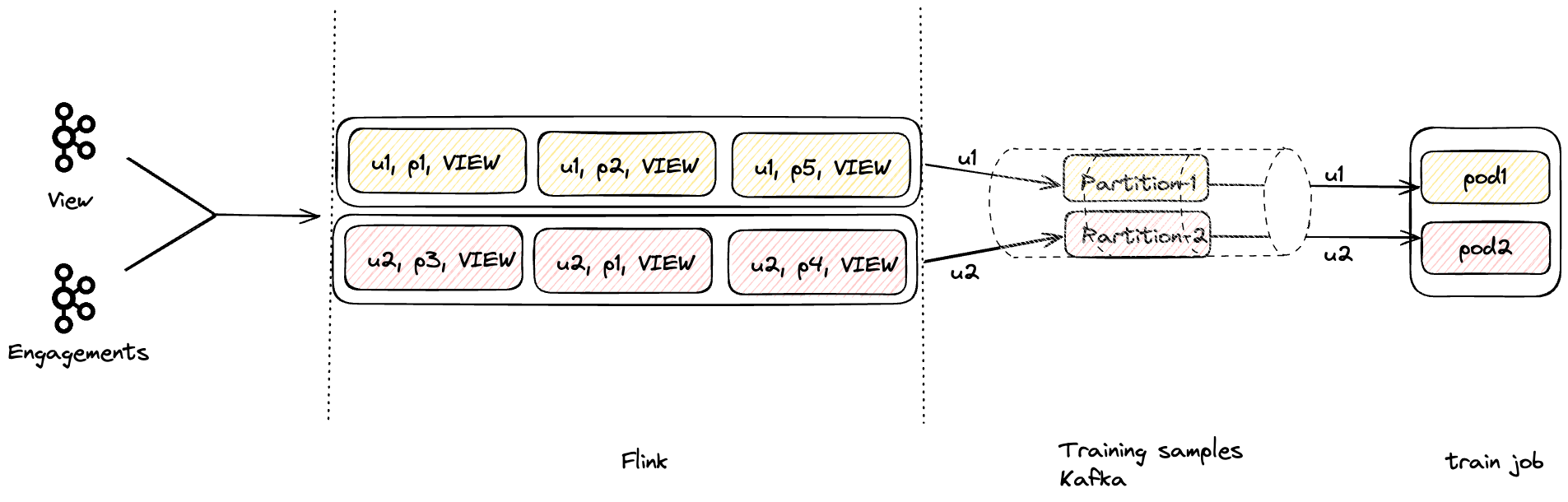}
    \vspace{-4mm}
\end{minipage}
% \hfill%
% \vspace{-10pt}
\caption{How pod contention can be solved using intrinsic properties of Kafka and Flink}
\label{fig:remove_redis}
% \vspace{-15pt}
\end{figure}

\subsubsection{Remove Redis} \label{flink_redis}
Finally, we can eliminate Redis using Keyed Streams in Flink and Topic partitioning in Kafka. In the example shown in Fig \ref{fig:remove_redis}, Flink processes events for users \textit{u}1 and \textit{u2} across distinct keyed streams. Flink directs these streams to separate partitions in Kafka, and the pods of the consumer job subsequently consume them. The crucial insight here lies in pre-allocating the partitions in Kafka and the pods in the training job. Without this pre-allocation, events for the same user could be assigned to different partitions during rebalancing, affecting the training. As seen, \textit{partition1} and \textit{pod1} will process all interactions by \textit{user1} while \textit{partition2} and \textit{pod2} will process all interactions by \textit{user2}.

\subsubsection{Other Optimisations} \label{compression}
Data compression is an effective strategy to reduce throughput \cite{datacompression, lz4}. We further optimized our pipeline by transitioning from JSON to Avro schema and employing \href{[https://en.wikipedia.org/wiki/LZ4_(compression_algorithm)](https://en.wikipedia.org/wiki/LZ4_(compression_algorithm))}{\textbf{LZ4}} compression, resulting in an 85\% reduction in the throughput.

\subsection{Analysing the Trade-Offs}
\subsubsection{At-least Once Delivery}
While exactly-once delivery is ideal for accuracy, it is complex to implement, and the additional complexity does not offset the gains. We implement a de-duplication layer in the training job, effectively making it an idempotent sink. So, at-least-once delivery in addition to an idempotent sink is a much more practical choice for us.

% \begin{figure}[!t]
% \begin{minipage}[b]{\linewidth}
%     \includegraphics[width=\textwidth]{figs/event_time.png}
%     \vspace{-4mm}
% \end{minipage}
% % \hfill%
% % \vspace{-10pt}
% \caption{Event1 is processed 5 hours later than when it actually happened, while Event2 is processed with a lag of few seconds.}
% \label{fig:event_time}
% % \vspace{-15pt}
% \end{figure}

\subsubsection{Event \& Processing Time}
\begin{figure}[!t]
\begin{minipage}[b]{\linewidth}
    \includegraphics[width=\textwidth]{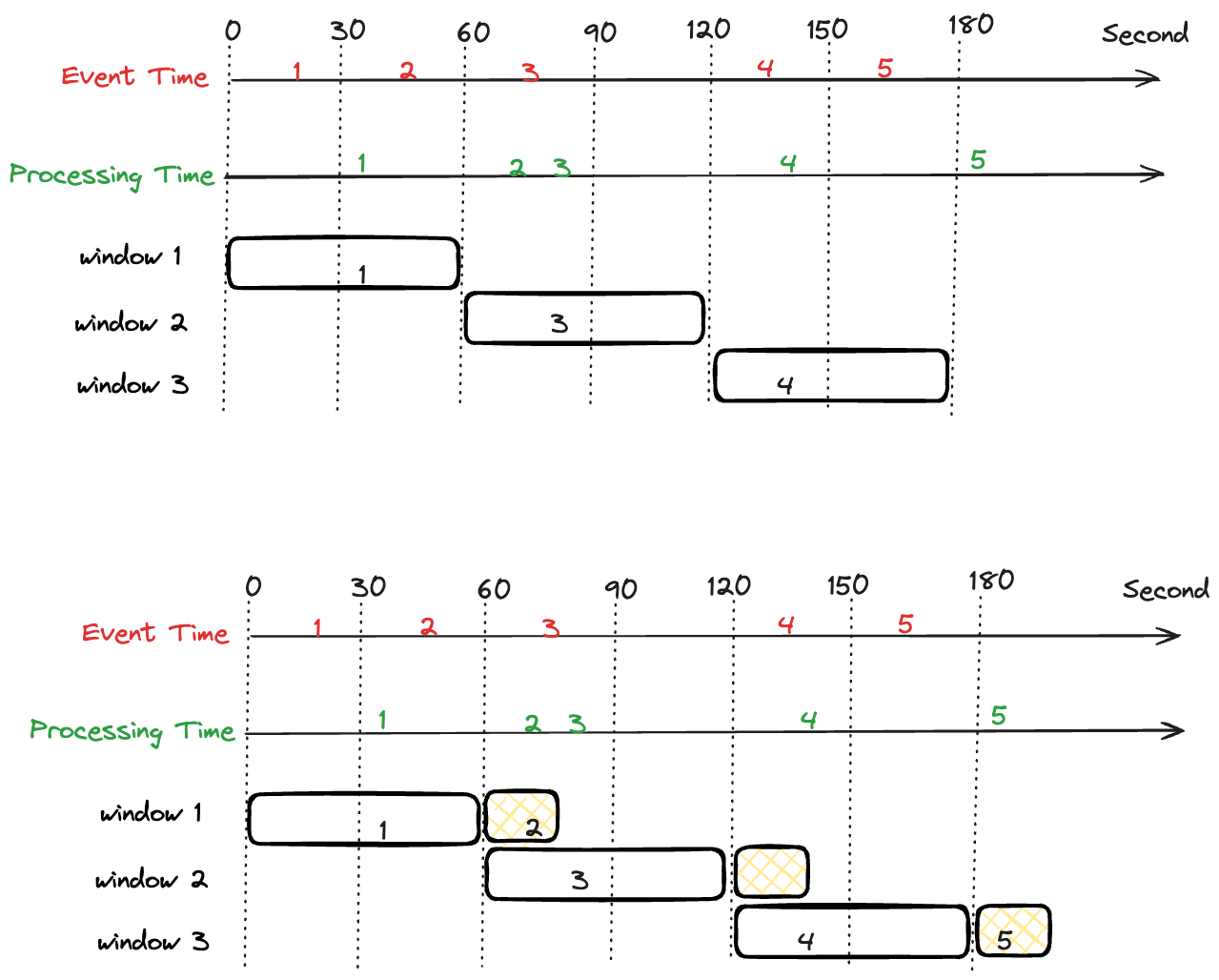}
    \vspace{-4mm}
\end{minipage}
% \hfill%
% \vspace{-10pt}
\caption{Demonstration of watermark in event processing.}
\label{fig:watermark_demo}
% \vspace{-15pt}
\end{figure}

We assess the tradeoffs between utilizing event time and processing time. Event time is when an event occurred, while processing time is when Flink starts processing the event. Their lag can be significant due to network delays and asynchronous environments, such as data transmission via message queues. The example in Fig \ref{fig:watermark_demo} shows the tradeoffs between event and processing time. Event time leads to more accurate results, but implementing it is more challenging, and we must deal with ``late events''. 

\subsubsection{Using Watermarks}
Let us say we have tumbling windows of size 1 min each. In Fig \ref{fig:watermark_demo}, window 1 misses event 2, and window 4 misses event 5. So, how do we solve this? We can use a watermark \cite{watermark, watermark_sysdesign} to extend our windows by an additional 15 sec. However, watermarks only assist with slightly delayed events, not those with significant delays. It is a tradeoff between latency and accuracy: extending the window boosts accuracy by capturing late events but adds latency \textit{t} to the system, and vice versa.

\section{Results \& Observations}

\begin{table}[h!]
\centering
\resizebox{\columnwidth}{!}{
\begin{tabular}{|c|c|c|>{\centering\arraybackslash}p{3cm}|}
\hline
\textbf{Components} & \textbf{Approach 1} & \textbf{Approach 2} & \textbf{Relative Cost Savings} \newline \textbf{(2 vs 1)} \\ \hline
Messaging System & PubSub & Kafka & \textbf{55\%} \\ \hline
Event Joiner & Golang Job + Memcache & Flink & \textbf{52\%} \\ \hline
Redis & \checkmark & - & \textbf{100\%} \\ \hline
Schema & Json & Avro & \multirow{2}{*}{\textbf{85\%}} \\ 
\cline{1-3}
Compression & - & LZ4 & \\ \hline
\end{tabular}
}
\vspace{1mm}
\caption{Comparison of Approaches and Cost Savings}
\end{table}

\subsection{Cost Comparison}
We compare the cost of the two systems described in Fig \ref{fig:system_comparison} by breaking it at the component level.
\subsubsection{PubSub vs.\ Kafka}
We compare the cost difference between two setups at a peak throughput of \textit{200 MBPS}, with a message retention period of 3 days for both, translating to around 10TiB of data daily. The Pub/Sub configuration uses a single topic with a single subscription, while the Kafka setup includes seven brokers. Currently, the Kafka setup is regional, but switching to a zonal setup could reduce costs by approximately 25\% due to lower inter-zone egress charges. However, this change would also result in reduced availability. For a similar setup, we observe the cost of Kafka as \textbf{55\%} lesser than PubSub. We can extrapolate the numbers linearly as there are five such topics, adjusting the throughput.

\subsubsection{Flink vs.\ Consumer Job}
The event joiner in Fig \ref{fig:rt_old} is a job written in Golang that consumes the events and joins them with real-time labels via Memcache. The Flink consumer does the same task in Fig \ref{fig:rt_new}. A GCS bucket is required to store checkpoints for the Flink job. The combined cost of Flink is \textbf{52\%} lesser than the Golang job.

\subsubsection{Redis and Memcache}
The cost of Memcache goes away with Flink. As described in Section \ref{flink_redis}, Flink solves pod contention using intrinsic properties. So, with Flink in place, Redis's cost also goes away.

\subsubsection{Data Compression}
As mentioned in \ref{compression}, using Avro schema and LZ4 compression resulted in an 85\% reduction in throughput, leading to similar savings in data ingestion cost.

\subsection{Performance or Latency}
Both the systems can be scaled horizontally to adjust according to the incoming traffic. The consumers in Approach 1 scale according to the number of unacked messages or oldest unacked message age in PubSub, along with CPU \& memory utilization. Flink also allows dynamic adjustment of resources based on workload, helping to optimize performance and cost. However, it requires re-partitioning in the Kafka topic associated with the job, the implementation of which is slightly complex. Currently, we provision Flink to handle maximum traffic and are exploring autoscaling as future work. Both setups ensure comparable performance after tuning, resulting in similar latency.

\subsection{Data Validation}
The switch from Approach 1 to Approach 2 was verified by setting up a system where the same events were duplicated and processed through both pipelines. Initially, we directed 0.1\% of the traffic through each pipeline to separate storage tables and compared the tables for accuracy. After confirming correctness (match rate \& schema), we gradually increased the traffic to 100\% before transitioning to production. Finally, we decommissioned the old pipeline.

\section{Conclusion \& Future Work}
This paper explores the implications of design choices for processing streaming events to generate training samples for machine learning models. We address key challenges such as real-time label joining, handling out-of-order events, and managing concurrent updates. By comparing trade-offs, we illustrate how to make informed design decisions for specific use cases. We further see how we can leverage the intrinsic properties of frameworks and platforms to simplify the system and make it more cost-effective. A one-line takeaway is that our decisions must consider cost, correctness, latency, maintainability, and the trade-offs we are prepared to accept.

The subsequent steps for this project involve implementing autoscaling for the Flink job. We also plan to minimize the source topic count by consolidating them under one topic, which will enhance system management and help further decrease costs. 

\section{Acknowledgements}
We sincerely thank Arya Ketan for his guidance on design choices and Shubham Dhal for his dedicated assistance with the implementation. Their contributions greatly improved the quality of this project.

\bibliographystyle{ACM-Reference-Format}
 \balance
\bibliography{sample-base}

\end{document}